\documentclass[prl,aps,preprint,amsmath,amssymb,showpacs]{revtex4}
\usepackage{graphicx}
\usepackage{epsfig}
\include{graphics}

\begin{document}

\title{Intermittent dynamics, strong correlations, and bit-error-rate in 
multichannel optical fiber communication systems}

\author{Avner Peleg}

\affiliation{Arizona Center for  Mathematical Sciences, University of
Arizona, Tucson, Arizona 85721, USA}
%\date{\today}

\begin{abstract}
We investigate the effects of delayed Raman response on pulse dynamics 
in massive multichannel optical fiber communication systems. 
Taking into account the stochastic 
nature of pulse sequences in different frequency channels 
and the Raman induced energy exchange in pulse collisions 
we show that the pulse parameters exhibit intermittent dynamic behavior,
and that the pulse amplitudes exhibit relatively strong and 
long-range correlations. Moreover, we find that 
the Raman-induced cross frequency shift is the main 
intermittency-related mechanism leading to bit pattern deterioration 
and evaluate the bit-error-rate of the system.  
\end{abstract}


\newpage

\maketitle
%\begin{multicols}{2}
The dynamic evolution of coherent patterns in the presence of noise 
and nonlinearities is a rich and complex subject that is of major 
importance in many areas of physics. Fiber optics communication 
systems, which employ optical pulses to represent bits of information, 
serve as an excellent example for systems where noise and nonlinear 
effects can have an important role in the dynamics of 
coherent patterns \cite{Agrawal2001}. 
It is known that the dynamics of the parameters characterizing the 
pulses in optical fiber transmission systems can exhibit non-Gaussian 
statistics \cite{Menyuk95,Falkovich2004,Turitsyn2005,Moore2005}. Yet, it is 
commonly believed that the statistics of the optical pulse parameters 
is very different from the intermittent statistics encountered in 
strong nonlinear phenomena such as turbulence and chaotic flow. 
(For a review of intermittency in the context of 
turbulent flow, see Ref. \cite{Frisch95}).  
In this Letter we present results that contrast this common belief 
and show that the parameters of optical pulses can exhibit 
intermittent dynamic behavior in massive multichannel transmission 
systems. Furthermore, we demonstrate that this intermittent 
dynamics can have important practical consequences by leading to 
relatively large values of the bit-error-rate (BER) characterizing the 
system performance.

We consider conventional optical solitons as an example for
the pulses carrying the information and focus attention on the 
effects of delayed Raman response on the propagation. The main effect 
of delayed Raman response on single-soliton propagation is the 
self frequency shift, which is due to energy transfer from 
higher frequency components of the pulse to its lower frequency 
components \cite{Mitschke86,Gordon86}. The main effect of a 
single two-soliton collision in the presence 
of delayed Raman response is an energy exchange between the 
colliding pulses (Raman induced cross talk), which leads to  
a change in their amplitudes \cite{Chi89,Malomed91,Agrawal96,
Kumar98,Kaup99}. In addition, the frequencies of the two solitons also
change as a result of the collision (Raman induced cross frequency
shift) \cite{Chi89,Agrawal96,Kumar98,Kaup99}.

The combined effect of Raman scattering and randomness of soliton 
sequences in multichannel transmission systems was considered in 
Refs. \cite{P2004,CP2005}, where it was found that the soliton 
amplitude has a lognormal distribution. It was also shown that the
distribution of the cross frequency shift in a two-channel 
system is lognormal and that the first two normalized moments of the 
self frequency shift grow exponentially with increasing distance. 
Even though these studies implied intermittent 
dynamic behavior for the soliton amplitude, it was not clear 
whether the other soliton parameters exhibit similar 
dynamic behavior in a general multichannel system. Moreover, 
the effect of the amplitude intermittent behavior on the 
main measure of system performance, the BER, was not addressed at all.
In this Letter we study in detail the intermittent character of 
soliton dynamics. We show that the normalized $n$th 
moments of the self and cross frequency shifts are 
exponentially increasing with both propagation distance 
$z$ and $n^{2}$, i.e., the self and cross frequency shifts exhibit 
intermittent dynamic behavior. We find that the $n$th order 
equal-$z$ amplitude correlation functions have similar 
dependence on $z$ and $n^{2}$ with a typical correlation time 
which is much larger than the time slot width, thus showing  
that the system exhibits relatively strong and long-range correlations.   
Furthermore, we find that the cross-frequency shift is the main 
intermittency-related mechanism leading to an increase of the BER, 
and calculate the $z$-dependence of the BER for different channels.

Propagation of pulses of light through an optical fiber in the presence 
of delayed Raman response is described by the following perturbed 
nonlinear Schr\"odinger equation \cite{Agrawal2001}:
\begin{eqnarray}
i\partial_z\Psi+\partial_t^2\Psi+2|\Psi|^2\Psi=
-\epsilon_{R}\Psi\partial_t|\Psi|^{2}.
\label{inter1}
\end{eqnarray}
In Eq. (\ref{inter1}) $\Psi$ is the envelope of the electric field, 
$z$ is the position along the fiber, $t$ is time in the 
retarded reference frame and the term 
$-\epsilon_{R}\Psi\partial_t|\Psi|^{2}$ accounts for 
the effect of delayed Raman response \cite{dimensions}.
When $\epsilon_{R}=0$, the single-soliton solution of Eq. (\ref{inter1})
in a given frequency channel $\beta$ is described by
$\Psi_{\beta}(t,z)\!=\!
\eta_{\beta}\exp(i\chi_{\beta})\cosh^{-1}(x_{\beta})$, where
$x_{\beta}=\eta_{\beta}\left(t-y_{\beta}-2\beta z\right)$
and $\chi_{\beta}=\alpha_{\beta}+\beta(t-y_{\beta})+
\left(\eta_{\beta}^2-\beta^{2}\right)z$, 
and $\alpha_{\beta},\eta_{\beta}$ and $y_{\beta}$ are the soliton
phase, amplitude, and position, respectively.

Consider a single collision between a soliton 
in the reference channel ($\beta=0$) and a soliton in the $\beta$
channel. We assume that $|\beta|\gg 1$, $\epsilon_{R}\ll 1$ and 
$\epsilon_{R}\ll 1/|\beta|$, which is the typical situation in
current multichannel transmission systems. Focusing attention on
changes in the parameters of the reference channel soliton, for example,  
one finds that the most important effect of the collision is an 
$O(\epsilon_{R})$ change in the soliton amplitude 
\cite{Malomed91,Kumar98}: 
\begin{eqnarray}
\Delta\eta_{0}=
2\eta_{0}\eta_{\beta}\mbox{sgn}(\beta)\epsilon_{R}.
\label{inter2}
\end{eqnarray}
The main effect of the collision 
in order $\epsilon_{R}/\beta$ is a frequency shift given 
by \cite{Kumar98}:
$\Delta\beta_{0}=-(8\eta_{0}^{2}\eta_{\beta}\epsilon_{R})/(3|\beta|)$. 
Since $\epsilon_{R}\ll 1/|\beta|\ll 1$, we neglect effects of order 
$\epsilon_{R}^{2}$ and higher.

We now describe propagation of a soliton in the reference channel undergoing 
many collisions with solitons from all other frequency channels in a system
with $2N+1$ channels. We assume that the amplitudes 
of the latter solitons are all equal to 1.  
The stochastic character of soliton sequences in different 
channels is taken into account by defining discrete random
variables $\zeta_{ij}$, which describe the occupation state of the
$j$th time slot in the $i$th channel: $\zeta_{ij}=1$ if the 
slot is occupied and 0 otherwise. Therefore, the $n$th 
moment of $\zeta_{ij}$ satisfies  
$\langle\zeta_{ij}^{n}\rangle=s$, where $s$ is the 
average fraction of occupied time slots, assumed to be the same
in all channels. We also assume that the occupation states of 
different time slots are uncorrelated: 
$\langle\zeta_{ij}\zeta_{i'j'}\rangle=s^{2}$ if $i\ne i'$ and 
$j\ne j'$. We denote by $\Delta\beta$ the frequency   
difference between neighboring channels and by $T$ the time slot width. 
We assume that the change in $\eta_{0}$ over the
interval $\Delta z_{c}^{(1)}=T/(2\Delta\beta)$, 
traveled by the reference channel soliton while passing 
two successive time slots in the nearby channels, is small.
Using Eq. (\ref{inter2}) and summing over all collisions occurring 
at $\Delta z_{c}^{(1)}$ we arrive at
\begin{eqnarray} &&
\!\!\!\!\!\left.\frac{\Delta\eta_{0}}
{\Delta z_{c}^{(1)}}\right|_{z=z_{k}}=
\frac{2\epsilon_{R}\eta_{0}(z_{k-1})}{\Delta z_{c}^{(1)}}
\sum_{i\ne 0}\mbox{sgn}(\beta_{i})
\!\!\!\!\!\!\sum_{j=(k-1)i+1}^{ki}\!\!\!\!\!\zeta_{ij}, 
\label{inter11}
\end{eqnarray}         
where $k-1$ and $k$ are the indexes of the two successive time slots 
in the $i=-1$ channel, $z_{k}-z_{k-1}=\Delta z_{c}^{(1)}$, and the 
outside sum is from $-N$ to $N$. We decompose the disorder 
$\zeta_{ij}$ into an average part and a 
fluctuating part: $\zeta_{ij}=s+\tilde\zeta_{ij}$,
where $\langle\tilde\zeta_{ij}\rangle=0$, 
$\langle\tilde\zeta_{ij}\tilde\zeta_{i'j'}\rangle=s(1-s)
\delta_{ii'}\delta_{jj'}$, and $\delta_{ii'}$ is the 
Kronecker delta function. Substituting $\zeta_{ij}=s+\tilde\zeta_{ij}$ 
into Eq. (\ref{inter11}) and going to the continuum limit we obtain
\begin{eqnarray} &&
\!\!\!\!\!\!\!\frac{1}{\eta_{0}}\frac{d\eta_{0}}{d z}=
\frac{4s\epsilon_{R}\Delta\beta}{T}
\sum_{i\ne 0}\mbox{sgn}(\beta_{i})|i|+2\epsilon_{R}\xi^{(0)}(z;N), 
\label{inter12}
\end{eqnarray}     
where the continuous disorder field $\xi^{(0)}(z;N)$ is
\begin{eqnarray} &&
\xi^{(0)}(z;N)=
\frac{1}{\Delta z_{c}^{(1)}}
\sum_{i\ne 0}\mbox{sgn}(\beta_{i})
\sum_{j=(k-1)i+1}^{ki}\!\!\!\!\!\tilde\zeta_{ij}.
\label{inter13}
\end{eqnarray} 
Using Eq. (\ref{inter13}) one can show that 
$\langle\xi^{(0)}(z;N)\rangle=0$ and 
$\langle\xi^{(0)}(z;N)\xi^{(0)}(z';N)\rangle=D^{(0)}_{N}\delta(z-z')$, 
where $D^{(0)}_{N}=N(N+1)D_{2}$, $D_{2}=2\Delta\beta s(1-s)T^{-1}$,   
and $\delta(z)$ is the Dirac delta function. Notice that the first 
term on the right hand side of Eq. (\ref{inter11}) is zero due to
symmetry. Integrating both sides of Eq. (\ref{inter11}) over $z$ we
obtain
\begin{eqnarray} &&
\eta_{0}(z)=\exp\left[2\epsilon_{R}x^{(0)}(z;N)\right],
\label{inter15}
\end{eqnarray}  
where $x^{(0)}(z;N)=\int_{0}^{z'}\mbox{d}z'\,\xi^{(0)}(z';N)$ and 
we assumed $\eta_{0}(0)=1$. According to the central limit theorem 
$x^{(0)}(z;N)$ is a Gaussian random variable with 
$\langle x^{(0)}(z;N)\rangle=0$ and 
$\langle x^{(0)2}(z;N)\rangle=D^{(0)}_{N}z$. 
As a result, the distribution of the soliton amplitude is lognormal
\begin{eqnarray} &&
\!\!\!\!\!F(\eta_{0})=(\pi {{\cal D}_{N}^{(0)}})^{-1/2}\eta_{0}^{-1}
\exp\left[-\ln^{2}\left(\eta_{0}\right)/{{\cal D}_{N}^{(0)}}\right]
\,,
\label{inter16}
\end{eqnarray}
where ${\cal D}_{N}^{(0)}=8D^{(0)}_{N}\epsilon_{R}^{2}z$.    
The normalized $n$th moment of $F(\eta_{0})$ satisfies
$\langle\eta_{0}^{n}(z)\rangle/\langle\eta_{0}(z)\rangle^{n}=
\exp\left[2n(n-1)D^{(0)}_{N}\epsilon_{R}^{2}z\right]$, 
from which it follows that the amplitude dynamics is 
intermittent.

For a soliton in the $i$th channel, 
the amplitude dynamics is given by 
\begin{eqnarray} &&
\frac{1}{\eta_{i}}\frac{d\eta_{i}}{d z}=
\frac{4s\epsilon_{R}\Delta\beta}{T}
\sum_{i'\ne i}\mbox{sgn}(\beta_{i'}-\beta_{i})
|i'-i|
 \nonumber \\ &&
+2\epsilon_{R}\xi^{(i)}(z;N), 
\label{inter17}
\end{eqnarray}        
where 
\begin{eqnarray} &&
\!\!\!\!\!\xi^{(i)}(z_{k};N)=
\frac{1}{\Delta z_{c}^{(1)}}
\sum_{i'\ne i}\mbox{sgn}(\beta_{i'}-\beta_{i})
\!\!\!\!\!\!\!\!\sum_{j=(k-1)(i'-i)}^{k(i'-i)}
\!\!\!\!\!\tilde\zeta_{i'j'}.
\label{inter18}
\end{eqnarray} 
From Eq. (\ref{inter18}) it follows that 
$\langle\xi^{(i)}(z;N)\rangle=0$ and 
$\langle\xi^{(i)}(z;N)\xi^{(i)}(z';N)\rangle=
D^{(i)}_{N}\delta(z-z')$, where 
$D^{(i)}_{N}=[N(N+1)+i^{2}]D_{2}$. Therefore, the disorder strength 
is different for different channels.  To solve Eq. (\ref{inter17}) 
we substitute $\eta_{i}(z)=\eta_{i}^{(d)}(z)\eta_{i}^{(f)}(z)$, 
where $\eta_{i}^{(d)}$ and $\eta_{i}^{(f)}$ represent the drift and 
fluctuating contributions due to the first and second terms 
on the right hand side of the equation, respectively. Assuming that 
the first term is compensated by appropriately
adjusting the amplifiers gain, $\eta_{i}^{(d)}(z)=1$ and 
$\eta_{i}(z)=\eta_{i}^{(f)}(z)$. As a result, the statistics of 
$\eta_{i}$ is described by the lognormal distribution (\ref{inter16}), 
with ${\cal D}_{N}^{(i)}=8D^{(i)}_{N}\epsilon_{R}^{2}z$ replacing
${\cal D}^{(0)}_{N}$.

\begin{figure}[ht]
\centerline{\scalebox{0.75}{\includegraphics{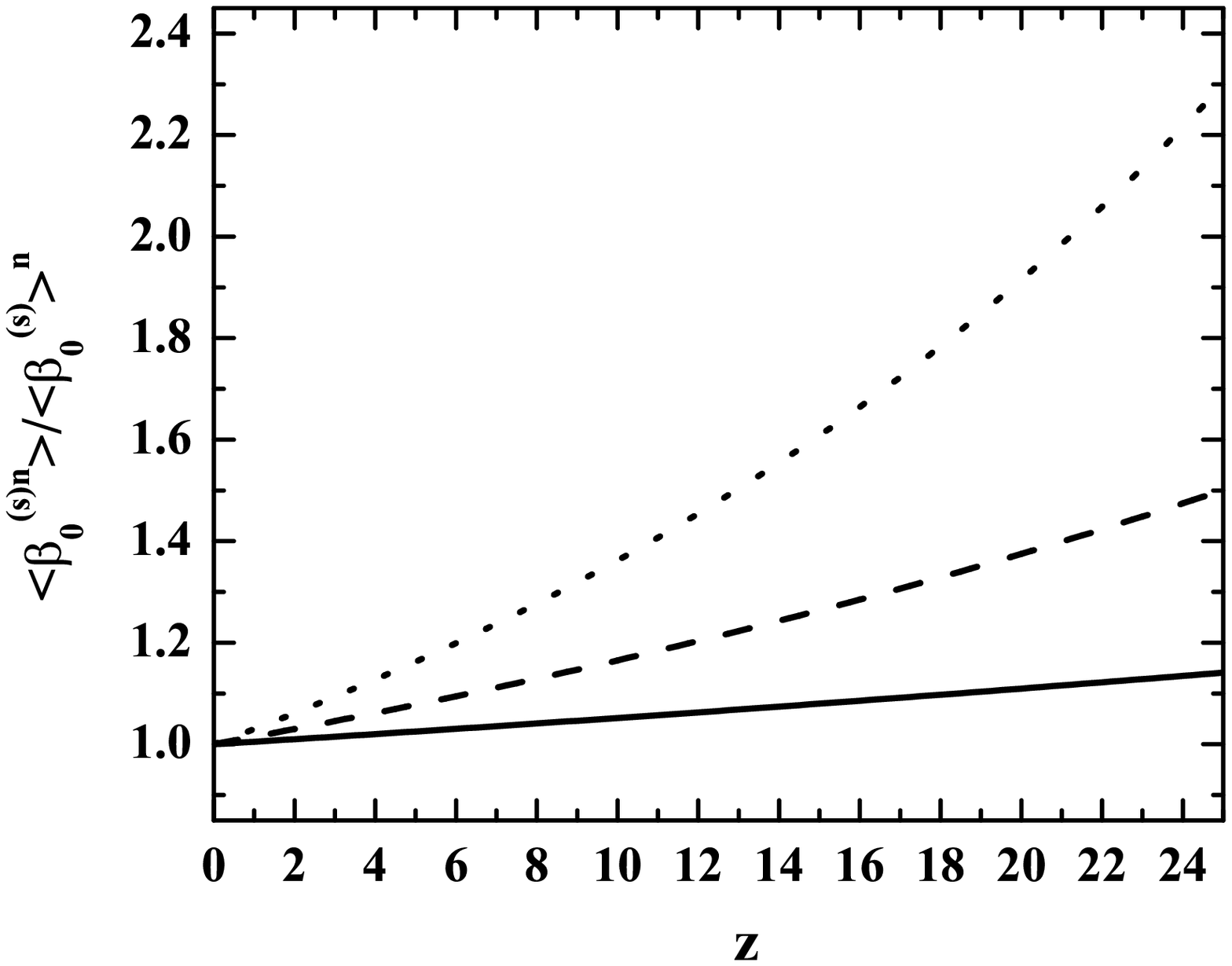}}} 
\caption{Normalized moments of the reference-channel soliton 
self frequency shift $\langle\beta_{0}^{(s)n}\rangle/
\langle\beta_{0}^{(s)}\rangle^{n}$ vs propagation 
distance $z$ for a 101-channel system operating at 10 Gbits/s per
channel. The solid, dashed, and dotted lines correspond to the $n=2,3,4$
normalized moments calculated by using Eq. (\ref{inter23}), 
respectively.}
\label{fig1}
\end{figure}

The dynamics of the Raman-induced self frequency shift for the 
reference channel soliton is given by
\begin{eqnarray} &&
\beta_{0}^{(s)}(z)=
-\frac{8\epsilon_{R}}{15}\int_{0}^{z}\mbox{d}z'
\eta_{0}^{4}(z'). 
\label{inter21}
\end{eqnarray}  
To show that Eq. (\ref{inter21}) leads to intermittent dynamics 
for $\beta_{0}^{(s)}$ we first calculate the $n$th moment:
\begin{eqnarray} &&
\langle\beta_{0}^{(s)n}(z)\rangle=
\left(-\frac{8\epsilon_{R}}{15}\right)^{n}n!
 \nonumber \\ &&
\times
\int_{0}^{z}\mbox{d}z_{1}\dots\int_{0}^{z_{n-1}}\mbox{d}z_{n}
\langle\eta_{0}^{4}(z_{1})\dots\eta_{0}^{4}(z_{n})\rangle, 
\label{inter22}
\end{eqnarray} 
where $0\le z_{n}\le\dots\le z_{1}\le z$.
Using Eq. (\ref{inter15}) and the fact that the integrals 
$\int_{z_{i-1}}^{z_{i}}\mbox{d}z'\,\xi^{(0)}(z';N)$ are Gaussian 
random variables that are independent for different $i$-values 
we obtain 
\begin{eqnarray} &&
\!\!\!\!\!\!\langle\beta_{0}^{(s)n}(z)\rangle=
\left(-\frac{8\epsilon_{R}}{15}\right)^{n}n!
 \nonumber \\ &&
\!\!\!\!\!\times\prod_{m=1}^{n}
\int_{0}^{z_{m-1}}\!\!\!\! \mbox{d}z_{m}
\exp\left[32D^{(0)}_{N}\epsilon_{R}^{2}(2m-1)z_{m}\right], 
\label{inter23}
\end{eqnarray} 
where $z_{0}=z$. Thus, $\langle\beta_{0}^{(s)n}(z)\rangle$ 
is given by a sum over exponential terms of the form 
$K_{m}\exp\left[32m^{2}D^{(0)}_{N}\epsilon_{R}^{2}z\right]$ 
where $0\le m\le n$ and $K_{m}$ are constants. 
To show intermittency it is sufficient to 
compare the leading term in the sum with the leading term in the 
expression for $\langle\beta_{0}^{(s)}(z)\rangle^{n}$. 
This calculation yields:    
\begin{eqnarray} &&
\frac{\langle\beta_{0}^{(s)n}(z)\rangle}
{\langle\beta_{0}^{(s)}(z)\rangle^{n}}\simeq
\frac{n!\exp\left[32n(n-1)D^{(0)}_{N}\epsilon_{R}^{2}z\right]}
{\prod_{m=1}^{n}\left[n^{2}-(m-1)^{2}\right]}. 
\label{inter24}
\end{eqnarray} 
Therefore, the leading term in the expression for the normalized $n$th 
moment of $\beta_{0}^{(s)}$ is exponentially increasing with both
$z$ and $n^{2}$. To illustrate this dynamic behavior we show the 
$z$-dependence of the $n=2,3,4$ normalized moments in Fig. \ref{fig1} 
for a multichannel transmission system with $N=50$, 
$\epsilon_{R}=3\times 10^{-4}$, $s=1/2$, $T=5$, and 
$\Delta\beta=10$. These parameters correspond to the 101-channel
system  operating at 10 Gbits/s per channel discussed in detail below.
One can see that the fourth moment increases much faster with increasing 
$z$ compared with the second and third moments.  
These results are of special importance since other contributions 
to changes in the soliton parameters are also coupled to the soliton 
amplitude via integrals over $z$ and thus follow the same statistics as
$\beta_{0}^{(s)}$. In particular, the $s$-dependent contribution 
to the cross frequency shift discussed below 
and the soliton's phase shift, which is
given by $\alpha_{0}(z)=2\int_{0}^{z}\mbox{d}z'\eta_{0}^{2}(z')$, 
exhibit similar intermittent dynamics.

The dynamics of the Raman-induced cross-frequency shift $\beta_{0}^{(c)}$ 
in a two-channel system was obtained in Refs. \cite{P2004,CP2005},  
where it was shown that $\beta_{0}^{(c)}$ is lognormally distributed. 
In a system with $2N+1$ channels one obtains by a 
procedure similar to the one used in deriving Eq. (\ref{inter11})
\begin{eqnarray} &&
\!\!\!\!\!\!\!\!\!\!\!\!\!\!\!\!\!\!\left.\frac{\Delta\beta_{0}^{(c)}}
{\Delta z_{c}^{(1)}}\right|_{z=z_{k}}\!\!\!\!\!\!\!\!\!\!\!=
-\frac{8\epsilon_{R}\eta_{0}^{2}(z_{k-1})}{3}
\sum_{i\ne 0}\frac{1}{|\beta_{i}|}
\!\sum_{j=(k-1)i+1}^{ki}
\!\!\!\!\frac{(s+\tilde\zeta_{ij})}{\Delta z_{c}^{(1)}}.
\label{inter31}
\end{eqnarray}       
Equation (\ref{inter31}) can be solved by decomposing $\beta_{0}^{(c)}$ 
into an $s$-dependent part $\beta_{0}^{(cd)}$ and an 
$s$-independent part $\beta_{0}^{(cf)}$. The $s$-dependent part 
is given by an integral of $\eta_{0}^{2}$ over $z$ 
and thus has similar statistics as $\beta_{0}^{(s)}$. 
To obtain the evolution of the $s$-independent part 
we decompose the total disorder field $\xi^{(0)}(z;N)$ 
into contributions $\xi_{i}^{(0)}(z)$ coming from different channels: 
$\xi^{(0)}(z;N)=\sum_{i\ne 0}\mbox{sgn}(\beta_{i})\xi_{i}^{(0)}(z)$, 
where $\xi_{i}^{(0)}(z)=\sum_{j}\tilde\zeta_{ij}/\Delta z_{c}^{(1)}$.
Hence, the fields $\xi_{i}^{(0)}$ satisfy $\langle\xi_{i}^{(0)}(z)\rangle=0$ 
and $\langle\xi_{i}^{(0)}(z)\xi_{i'}^{(0)}(z')\rangle=
|i|D_{2}\delta_{ii'}\delta(z-z')$. Substituting these relations into 
Eq. (\ref{inter31}), going to the continuum limit and integrating 
over $z$ we obtain
\begin{eqnarray} &&
\!\!\!\!\!\beta_{0}^{(cf)}(z)= 
-\frac{2}{3}\eta_{0}^{2}(z)
\sum_{i\ne 0}\frac{\mbox{sgn}(\beta_{i})}{|\beta_{i}|}
\left[1-\mu_i^{-1}(z)\right],
\label{inter33}
\end{eqnarray} 
where $\mu_i(z)=\exp\left[4\mbox{sgn}(\beta_{i})\epsilon_{R}
x_{i}^{(0)}(z)\right]$, and 
$x_{i}^{(0)}(z)=\int_{0}^{z}\mbox{d}z'\,\xi_{i}^{(0)}(z')$.
Therefore, $\beta_{0}^{(cf)}(z)$ is a product of a lognormal 
variable and a sum over independent lognormal variables. 
Since the terms in the sum on the right hand side 
of Eq. (\ref{inter33}) decrease with frequency as $1/|\beta_{i}|$ 
it is sufficient to consider only contributions from a few  
neighboring channels. For a three-channel system, for example, 
Eq. (\ref{inter33}) simplifies to
\begin{eqnarray} &&
\beta_{0}^{(cf)}(z)= 
\frac{2}{3\Delta\beta}
\left[\mu_{-1}(z)-\mu_{1}(z)\right].
\label{inter34}
\end{eqnarray}  
Using Eq. (\ref{inter34}) one finds that the leading term 
in the expression for the normalized $2n$th moment of $\beta_{0}^{(cf)}$ 
is $\exp\left[32n(n-1)D^{(0)}_{N}\epsilon_{R}^{2}z\right]
/2^{n}$, which is exponentially growing with $z$. 
Hence, the cross-frequency shift in the 
three-channel system exhibits  intermittent dynamic behavior, 
even though it is not lognormally distributed as in the two-channel case.
 
To gain further insight into the intermittent dynamic behavior 
exhibited by the solitons we calculate the $n$th order 
equal-distance amplitude correlation functions, which measure
correlation between amplitudes of solitons from different 
time slots in the same channel. Considering the reference channel 
we calculate 
\begin{eqnarray} &&
C_{0j}=\langle\left(\eta_{00}\eta_{0j}\right)^{n}\rangle/
(\langle\eta_{00}^{n}\rangle\langle\eta_{0j}^{n}\rangle)-1,
\label{inter41}
\end{eqnarray}   
where $\eta_{00}$ and $\eta_{0j}$ stand for the amplitudes of the 
solitons in the $0$th and $j$th time slots, respectively. 
We still assume that the amplitudes of solitons in other channels
are 1. Therefore, considering collisions with solitons from the 
$i$th channel, for example, the difference between the dynamics of 
the two solitons is due to the fact that the $0j$ soliton
experiences the disorder experienced by the $00$ soliton with 
a delay given by $\Delta z_{ji}=jT/(2\beta_{i})$. Using this 
fact and the decomposition of $\xi^{(0)}$ into the 
$\xi_{i}^{(0)}$, and assuming $j>N$, one can show that
$\tilde C_{0j}(z)=\eta_{00}(z)\eta_{0j}(z)$ is lognormally
distributed
\begin{eqnarray} &&
F(\tilde C_{0j})=
\frac{\exp\left\{-\ln^{2}\left(\tilde C_{0j}\right)/
[8\epsilon_{R}^{2}\tilde D_{N}^{(0)}(z)]\right\}}
{\left[8\pi\epsilon_{R}^{2}\tilde D_{N}^{(0)}(z)
\right]^{1/2}\tilde C_{0j}},
\label{inter45}
\end{eqnarray} 
where
\begin{eqnarray} &&
\tilde D_{N}^{(0)}(z)= 
4D_{2}\left[N(N+1)-\frac{1}{2}|i|(|i|-1)\right]z
 \nonumber \\ &&
-\frac{2D_{2}T(N-i+1)|j|}{\Delta\beta},
\label{inter44}
\end{eqnarray}  
for $\Delta z_{ji}< z< \Delta z_{j(i-1)}$. 
Consequently, the normalized $n$th order equal-distance amplitude 
correlation functions are given by
\begin{eqnarray} &&
\!\!\!\!\!\!\!C_{0j}^{(n)}(z)=\exp\left\{2n^{2}\epsilon_{R}^{2}
\left[\tilde D_{N}^{(0)}(z)-2D_{N}^{(0)}z\right]\right\}-1.
\label{inter46}
\end{eqnarray}            
In particular, for $z<\Delta z_{jN}$ $C_{0j}(z)=0$ since the two 
solitons are uncorrelated. During the transient 
$\Delta z_{jN}<z<\Delta z_{j1}$ the solitons become 
correlated due to the effective collision-induced disorder.    
For $z>\Delta z_{j1}$, i.e.,  after the transient, 
\begin{eqnarray} &&
\!\!\!\!\!\!\!\!\!\!\!\!\!\!\!
C_{0j}^{(n)}(z)\!=\exp\!\left[4n^{2}D_{N}^{(0)}\epsilon_{R}^{2}z
\!-\!8s(1-s)n^{2}N\epsilon_{R}^{2}|j|\right]\!\!-\!\!1.
\label{inter47}
\end{eqnarray}
Thus, after the transient the $n$th order correlation functions grow 
exponentially with both $n^{2}$ and $z$, in accordance with the  
intermittent behavior of the amplitude. Notice that $C_{0j}^{(n)}$ 
decays exponentially with $N|j|$, which is the total number of time 
slots in all other channels separating the two solitons. 
Using Eq. (\ref{inter47}) with $n=1$ one obtains $
j_{cor}=1/(2N\epsilon_{R}^{2})$ for the typical correlation number. 
For a 101-channel system operating at 10Gbits/s per channel, 
$j_{cor}\sim 10^{5}$, which is much larger than 
the number of successive bits that can be corrected 
by current error correction methods ($\sim 10^{3}$), 
and much smaller than the number of time slots that are 
in transmission in a given channel at any given 
time ($\sim 10^{8}$). Thus, this type of effective  
collision-induced disorder presents a challenge for conventional 
error correction methods.

We now relax the frozen disorder assumption and take into account 
the dynamics of soliton amplitudes in all channels. 
In this case Eq. (\ref{inter12}) is replaced by 
\begin{eqnarray} &&
\!\!\!\!\!\!\frac{1}{\eta_{0}}\frac{d\eta_{0}}{d z}=
\frac{4s\epsilon_{R}\Delta\beta}{T}
\sum_{i\ne 0}|i|\eta_{i}(z)\mbox{sgn}(\beta_{i})
 \nonumber \\ &&
+2\epsilon_{R}\hat\xi^{(0)}(z;N), 
\label{inter52}
\end{eqnarray}    
where $\eta_{i}(z)=\exp\left[2\epsilon_{R}
\int_{0}^{z}\mbox{d}z'\,\xi^{(i)}(z';N)\right]$
and 
\begin{eqnarray} &&
\!\!\!\!\!\!\!\!\!\!\hat\xi^{(0)}(z;N)=
\frac{1}{\Delta z_{c}^{(1)}}
\sum_{i\ne 0}\eta_{i}(z)\mbox{sgn}(\beta_{i})
\!\!\!\sum_{j=(k-1)i+1}^{ki}\!\!\!\tilde\zeta_{ij}.
\label{inter53}
\end{eqnarray} 
Expressing $\eta_{0}(z)$ as a product of an $s$-dependent and an 
$s$-independent parts: $\eta_{0}(z)=\eta_{0}^{(d)}(z)\eta_{0}^{(f)}(z)$, 
substituting into Eq. (\ref{inter52}), and integrating over $z$ we 
obtain
\begin{eqnarray} &&
\!\!\!\!\!\!\!\!\!\!\!\!\!\!\!\!\eta_{0}^{(d)}(z)=
\exp\left[\frac{4\Delta\beta s\epsilon_{R}}{T}
\sum_{i\ne 0}|i|\mbox{sgn}(\beta_{i})\int_{0}^{z}\mbox{d}z'\,
\eta_{i}(z')\right],
\label{inter54}
\end{eqnarray} 
and
\begin{eqnarray} &&
\eta_{0}^{(f)}(z)=\exp\left[2\epsilon_{R}\int_{0}^{z}\mbox{d}z'\,
\hat\xi^{(0)}(z';N)\right].
\label{inter55}
\end{eqnarray} 
It follows that  $\eta_{0}^{(d)}(z)$ is no longer deterministic. 
Moreover, since $\eta_{0}^{(d)}(z)$ is proportional to the exponent 
of the integral over $\eta_{i}(z)$, where $\eta_{i}(z)$ 
is lognormal, one can expect the departure of  
the $\eta_{0}^{(d)}$ statistics from Gaussian statistics to be stronger 
than lognormal. Consider now the statistics of $\eta_{0}^{(f)}$.
Using Eq. (\ref{inter53}) one can show that 
$\langle\hat\xi^{(0)}(z;N)\rangle=0$ and 
$\langle\hat\xi^{(0)}(z;N)\hat\xi^{(0)}(z';N)\rangle=
\hat D_{N}^{(0)}(z)\delta(z-z')$, where 
\begin{eqnarray} &&
\hat D_{N}^{(0)}(z)=D_{2}
\sum_{i\ne 0}|i|
\exp\left[8D_{N}^{(i)}\epsilon_{R}^{2}z\right],
\label{inter56}
\end{eqnarray} 
and we assumed
$\langle\eta_{i}(z)\eta_{i'}(z')
\tilde\zeta_{ij}\tilde\zeta_{i'j'}\rangle=
\langle\eta_{i}(z)\eta_{i'}(z')\rangle
\langle\tilde\zeta_{ij}\tilde\zeta_{i'j'}\rangle$. As a result, 
the $\eta_{0}^{(f)}$-distribution is the lognormal distribution 
given by Eq. (\ref{inter16}) with ${\cal D}_{N}^{(0)}$ replaced by
${\cal \hat D}_{N}^{(0)}(z)$, where   
\begin{eqnarray} &&
\!\!\!\!\!\!\!\!\!\!{\cal \hat D}_{N}^{(0)}(z)=D_{2}
\sum_{i\ne 0}\frac{|i|}{D_{N}^{(i)}}
\left[\exp\left(8D_{N}^{(i)}\epsilon_{R}^{2}z\right)-1\right].
\label{inter57}
\end{eqnarray}        
A direct consequence of Eq. (\ref{inter57}) is that the 
$n$th moments of the $\eta_{0}^{(f)}$-distribution are super-exponentially 
increasing with $z$, although the factors $8D_{N}^{(i)}\epsilon_{R}^{2}z$ 
are much smaller than 1.

From the practical point of view it is important to understand the 
influence of the intermittent dynamic behavior of the soliton 
parameters on the BER. The contribution of the collision-induced 
pulse decay to the BER was discussed in detail in previous 
works (see Ref. \cite{Tkach97} and references therein). 
Moreover, the small-$\eta$ tail of the lognormal distribution lies below 
the corresponding tail of the Gaussian distribution, whereas the 
large-$\eta$ lognormal tail lies above the corresponding Gaussian tail. 
As a result, strong effects due to deviations from Gaussian statistics 
are related to relatively large $\eta$-values. When the position 
dynamics or the frequency dynamics are coupled to the amplitude dynamics 
such large $\eta$-values can lead to significant increase in the BER due 
to walk-off of the soliton from its assigned time slot. 
Therefore, we focus our attention on contributions to the BER 
due to the large-$\eta$ lognormal tail. We consider a 101-channel 
system operating at 10Gbits/s per channel and emphasize that 
state-of-the-art experiments with dispersion-managed solitons 
demonstrated multichannel transmission with 109 channels 
at 10 Gbits/s per channel over a distance of 
$2\times 10^{4}$ km \cite{Mollenauer2003}.  
We use the following parameters, which are similar to the 
ones used in multichannel soliton transmission experiments \cite{MM98}. 
Assuming that $T=5$, $\Delta\beta=10$ and $s=1/2$, the pulse width 
is 20 ps, $\epsilon_{R}=3\times 10^{-4}$, the channel spacing is 
75 GHz, and $D_{2}=1$. Taking $\beta_{2}=-1\mbox{ps}^{2}/\mbox{km}$,  
the soliton-peak-power is $P_{0}=1.25$ mW. For these values 
the disorder strength is ${\cal D}^{(0)}_{50}(z)=1.8\times 10^{-3}z$ 
for the reference channel and 
${\cal D}^{(50)}_{50}(z)=3.6\times 10^{-3}z$  
for the two outermost channels. For $z=25$, corresponding to
transmission over $2\times 10^{4}$ km, 
${\cal D}^{(0)}_{50}(25)=0.046$ and  
${\cal D}^{(50)}_{50}(25)=0.091$.

\begin{figure}[ht]
\centerline{\scalebox{0.75}{\includegraphics{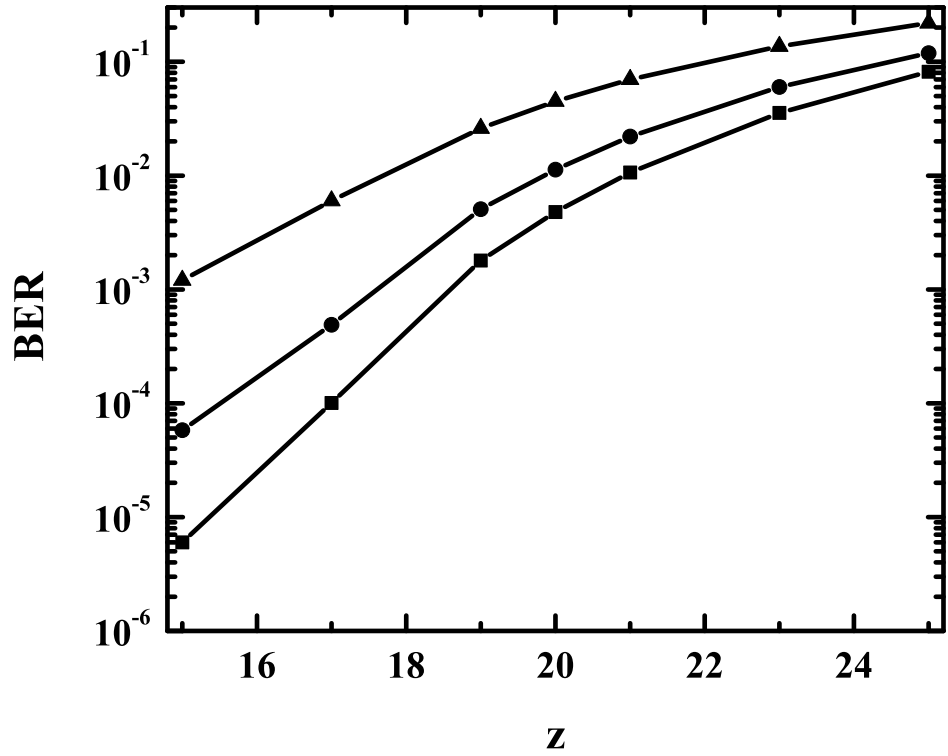}}} 
\caption{BER vs propagation distance $z$ for a 101-channel 
transmission system operating at 10 Gbits/s per channel. The 
squares, circles, and triangles represent 
the BER at channels $i=0$ (central), $i=25$, 
and $i=50$ (outermost), respectively.}
\label{fig2}
\end{figure}

For this system we evaluated the contributions to BER 
due to the Raman-induced cross frequency shift, 
the Raman-induced self frequency shift and the ``ideal'' 
component of the collision-induced position shift, i.e., 
the position shift due to soliton collisions in the absence of 
perturbations. The calculations show that the dominant 
contribution to the BER is due to the $s$-dependent part 
of the cross-frequency shift $\beta_{0}^{(cd)}$. 
The position shift induced by $\beta_{0}^{(cd)}$ 
is obtained by taking the continuum limit in Eq. (\ref{inter31})  
and integrating the $s$-dependent term twice with respect to $z$: 
\begin{eqnarray} &&
y_{0}^{(cd)}(z)= 
-\frac{64N\epsilon_{R}s}{3T}
\int_{0}^{z}\mbox{d}z'\int_{0}^{z'}\mbox{d}z''\eta_{0}^{2}(z'').
\label{inter62}
\end{eqnarray}
The position shift with a fixed amplitude $\eta_{0}(z)=1$ is
$\tilde y_{0}^{(cd)}(z)= -(32N\epsilon_{R}s z_{f}^{2})/(3T)$ and 
the relative position shift is 
$\Delta  y_{0}^{(cd)}(z)=y_{0}^{(cd)}(z)-\tilde y_{0}^{(cd)}(z)$. 
We assume that $\tilde y_{0}^{(cd)}$ can be compensated by 
employing filters. Therefore, the total energy of the 
soliton at a distance $z$ is 
\begin{eqnarray} &&
\!\!\!\!\!\!\!\!\!\!\!\!\!\!\!\!\!\!
I(\eta_{0},\Delta y_{0}^{(cd)})= 
\eta_{0}^{2}\int_{-T/2}^{T/2}\!\!\mbox{d}t
\cosh^{-2}[\eta_{0}(t-\Delta y_{0}^{(cd)})].
\label{inter66}
\end{eqnarray}  
Occupied time slots are considered to be in error, if 
$I(\eta_{0},\Delta y_{0}^{(cd)})\le I(z=0)/2\simeq1$. 
To estimate the BER we numerically integrate Eq. (\ref{inter62}) 
coupled to Eq. (\ref{inter15}) for different realizations of the
disorder $\xi^{(0)}(z,N)$ and calculate the fraction of errored 
occupied time slots. The BER in a generic channel $i\ne 0$ is 
calculated in a similar manner, where $\eta_{0}$ and $\xi^{(0)}(z,N)$  
are replaced by $\eta_{i}$ and $\xi^{(i)}(z,N)$, respectively.
Figure \ref{fig2} shows the $z$-dependence 
of the BER in channels $i=0$ (the reference channel), $i=25$, and $i=50$ 
(the outermost channel) for the aforementioned system. One can see that the 
BER in the reference channel increases from values smaller than $10^{-5}$ for 
$z<15$ ($x<1.2\times 10^{4}$ km) to about $8.2\times 10^{-2}$ at 
$z=25.0$ ($x=2\times 10^{4}$ km). Furthermore, for intermediate 
distances $15<z<20$, the BER value in the outermost channels 
can exceed that in the reference channel by several orders of magnitude, 
even though, the disorder strengths differ by only a factor of 2. 
This behavior presents another challenge to conventional error correction 
methods based on knowledge gained from single- or few-channel 
transmission systems.  

\begin{figure}[ht]
\centerline{\scalebox{0.75}{\includegraphics{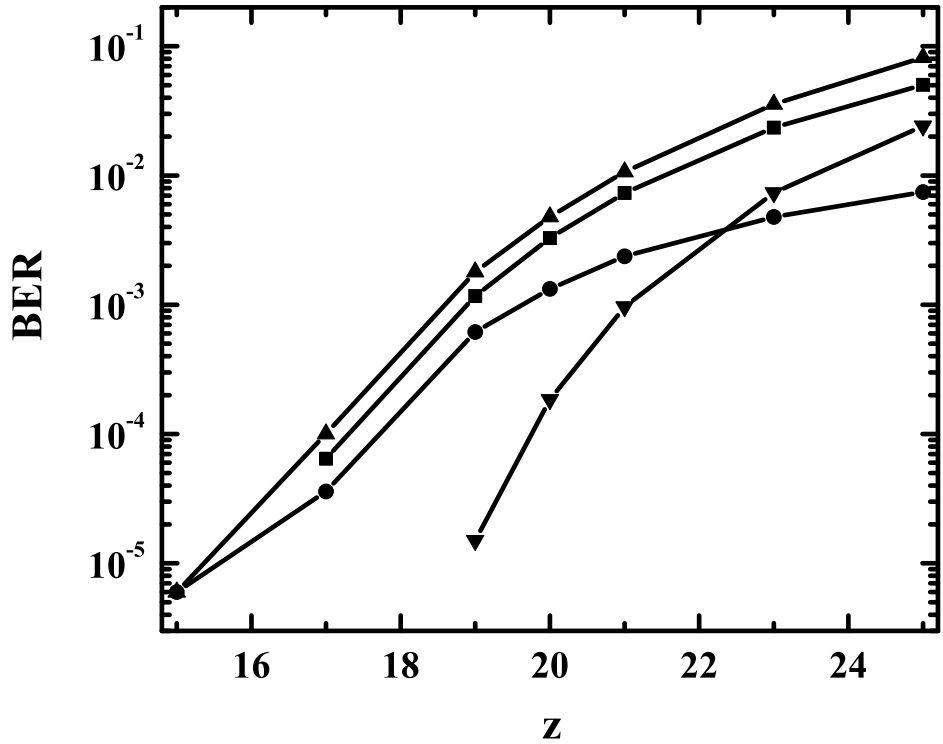}}} 
\caption{The $z$-dependence of different contributions to the BER for the 
reference channel in a 101-channel system operating at 10 Gbits/s 
per channel. The up triangles correspond to the total BER. The squares, 
circles, and down triangles correspond to contributions coming from the 
regions $\eta>1.0$ and $y<0$, $\eta<0.7$ and $y>0$, and 
$0.7<\eta<1.0$, respectively, in the $\eta_{0}-\Delta y_{0}^{(cd)}$ 
plane.}
\label{fig3}
\end{figure}

\begin{figure}[ht]
\centerline{\scalebox{0.75}{\includegraphics{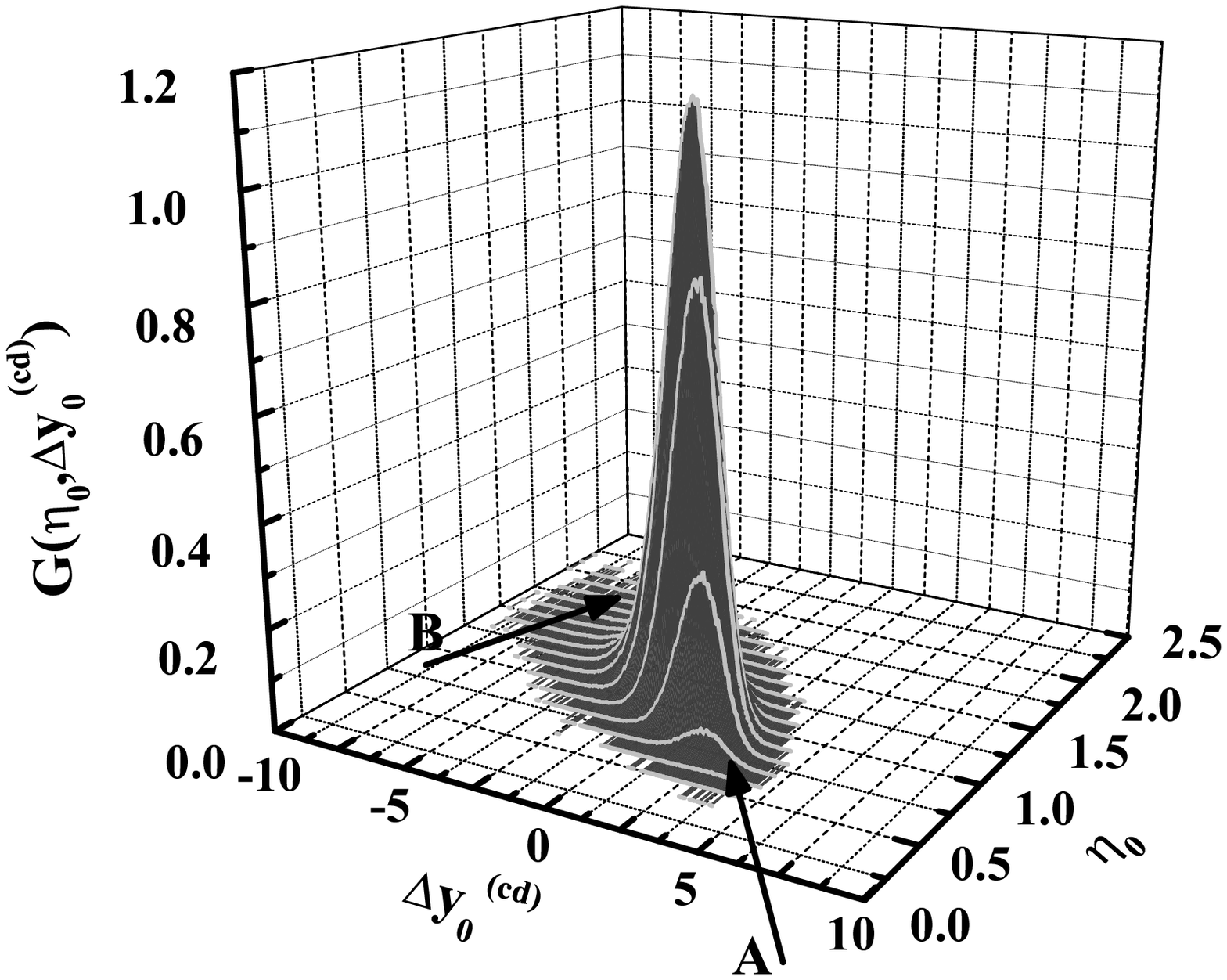}}} 
\caption{Mutual distribution function $G(\eta_{0},\Delta y_{0}^{(cd)})$ 
for a reference channel soliton at z=25 in a 101-channel system 
operating at 10 Gbits/s per channel. The arrows A and B show the domains
giving the main contributions to the BER.}
\label{fig4}
\end{figure}

To better understand error generation due to $\beta_{0}^{(cd)}$ 
we analyzed the $z$-dependence of contributions to the BER coming from 
different regions in the $\eta_{0}-\Delta y_{0}^{(cd)}$ 
plane. The results of this analysis are presented in Fig. (\ref{fig3}).
At $z=15$ the dominant contribution to the BER comes from the 
domain $\eta_{0}<0.7$ and $\Delta y_{0}^{(cd)}>0$, i.e., from 
decaying solitons with relatively large positive values of 
$\Delta y_{0}^{(cd)}$. For $z\ge 17$ the dominant contribution comes 
from the region $\eta_{0}>1.0$ and $\Delta y_{0}^{(cd)}<0$, 
which corresponds to solitons with relatively large amplitudes 
and large negative values of $\Delta y_{0}^{(cd)}$. The latter 
contribution is associated with the large-$\eta$ lognormal tail 
of the amplitude distribution. Figure \ref{fig4} shows the 
mutual distribution function $G(\eta_{0},\Delta y_{0}^{(cd)})$ 
at $z=25$ and the two domains giving the main contributions to the 
BER. It can be seen that this distribution is very different from the
one observed for single-channel soliton propagation 
in the presence of amplifier noise (see Fig. 1 in Ref. 
\cite{Falkovich2004}). While the latter 
distribution is approximately symmetric about $\Delta y_{0}=0$ and 
$\eta_{0}=1$, the former is strongly asymmetric with an extended tail 
in the large-$\eta_{0}$ and large-negative-$\Delta y_{0}^{(cd)}$ 
region. The strong asymmetric form of 
$G(\eta_{0},\Delta y_{0}^{(cd)})$ in our case is due to the strong coupling 
between the position dynamics and the amplitude dynamics and the 
lognormal statistics of the soliton amplitude. Thus, we find that 
amplitude dynamics plays a dominant role in error generation in 
massive multichannel optical fiber transmission systems, a situation 
which is very different from the one observed in single-channel transmission
systems \cite{Falkovich2004}.

In summary, we studied soliton propagation in massive multichannel 
optical fiber communication systems taking into account 
the effects of delayed Raman response and the random 
character of pulse sequences. We found that the soliton parameters
exhibit intermittent dynamic behavior and showed that 
the cross frequency shift is the main mechanism leading to 
bit pattern deterioration and to relatively large 
values of the bit-error-rate. We emphasize that similar 
dynamic behavior is expected in massive dispersion-managed 
multichannel transmission systems as well. 
In such systems the Raman-induced energy exchange in collisions 
will lead to lognormal statistics for the pulse amplitudes. In addition,
the frequency and position dynamics will be affected by a variety of 
amplitude-dependent perturbations due to Kerr nonlinearity. The coupling 
of the frequency and position dynamics to the amplitude dynamics will 
lead to intermittent dynamics of the pulse frequency and position and 
to relatively large values of the bit-error-rate.

\newpage

%\end{multicols}

\begin{thebibliography}{}
\bibitem{dimensions} The dimensionless $z$ in Eq. (\ref{inter1}) is
$z=(\beta_{2} x)/(2\tau_{0})$, where $x$ is the actual position,
$\tau_{0}$ is the soliton width, and $\beta_{2}$ is the second order
dispersion coefficient. The dimensionless retarded time is
$t=\tau/\tau_{0}$, where $\tau$ is the retarded time.  
The spectral width is $\nu_{0}=1/(\pi^{2}\tau_{0})$ and the channel spacing is
$\Delta\nu=(\pi\Delta\beta\nu_{0})/2$.  
$\Psi=E/\sqrt{P_{0}}$, where $E$ is the 
electric field and $P_{0}$ is the peak power. 
The dimensionless second order dispersion coefficient is 
$d=-1=\beta_{2}/(\gamma P_{0}\tau_{0}^{2})$, 
where $\gamma$ is the Kerr nonlinearity coefficient.
The dimensionless Raman coefficient is
$\epsilon_{R}=0.006/\tau_{0}$, where $\tau_{0}$ is in picoseconds.
\end{thebibliography}
\end{document}